\title{Confirmative laboratory tests and one example of forensic application of the probabilistic approach to the area of convergence in BPA}
\author{
        Francesco Camana\footnote{Italian State Police (\emph{Polizia di Stato}), Interregional forensic science office (\emph{G.I.P.S.}), Piazzetta Palatucci,~5 - 35123 Padova (ITALY)}~\footnote{\textit{E-mail}: francesco.camana@interno.it}~, Massimiliano Gori$^*$,\\ Nicola Gravina$^*$, Marco Quintarelli$^*$
}
\date{\today}

\documentclass[12pt]{article}
\usepackage{graphicx}
\usepackage{subfigure}
\usepackage{epstopdf}
\usepackage{capt-of}
\usepackage{cite}

\newcommand{\goodgap}{%
\hspace{\subfigtopskip}%
\hspace{\subfigbottomskip}}

\begin{document}
\maketitle

\begin{abstract}
One of the most important results in Bloodstain Pattern Analysis (BPA) is the determination of the area of convergence of blood-drop trajectories. This area is directly related to the point of origin of the projections and is often indicative of the point where the main action of a crime has occurred.
One of us has recently proposed a method to statistically characterize this area by mean of a probabilistic approach based on the uncertainties of the angles of impact of the stains in the pattern.
In our work we present some laboratory tests that confirm the validity of the method, returning good agreement between the empirical and the theoretical data. By comparing the results of different operators, we also show the robustness of the method, in that the results are independent of the analytical approach of the single experimenter.
Finally, we describe an example of application to a real forensic case.
\end{abstract}

\begin{scriptsize}
\end{scriptsize}

\section{Introduction}\label{Intro}
In the past few years, a growing interest has been registered in the field of the interpretation of the blood patterns at crime scenes. In particular, some of the most relevant advances regarded the methods for calculating the area of convergence and the point of origin of blood projections \cite{BHC,CH}. As a matter of fact, the possibility to detect the exact point of the scene where peculiar actions may have occurred has important forensic implications, both for crime scene reconstruction in itself and for the verification of the testimonies. 

Recent studies and forensic practice have produced a widely accepted procedure of identification of this point of origin in 3D, which follows these ordered steps: - careful selection and analysis of the bloodstains \cite{W,BSL,IB2}; - estimation of the angles of impact and their uncertainty \cite{WPDBR,P}; - projection of these angles on the horizontal plane and identification of the area of convergence of the projections \cite{IB,C}; - reconstruction of the ballistic trajectories in 3D \cite{BHC,CH,BSL,K,BKNAST,VG,CIF}.
Each of these phases has its own specific rules and methods, and we refer to the cited works and to \cite{BG,JE} for further details.

In this paper we want to focus our attention on the determination of the area of convergence. This is a key step for two main reasons: first of all the area of convergence, being the horizontal projection of the point of origin of blood, is directly linked to the position of the victim and/or the offender in the scene at a precise moment; second, this area is technically the region where the projected trajectories intersect, and therefore its position is also the starting point for determining the height of origin of projections.

One of the authors of this paper has recently developed a method for characterizing this area \cite{FC}; this procedure is simply embeddable in a forensic software and provides immediate and mathematically definite results. In this paper we show how some performed tests seem to confirm the validity of that analytical scheme, particularly referring to the statistical characterization and to the immediate visualization of the outcomes.

We finally describe one example of forensic application of the method, where the BPA was essential for the crime scene reconstruction and where the determination of the areas of convergence permitted a verification of the testimonies and, ultimately, the incrimination of the murderer.

\subsection{A recall of the probabilistic approach}
The method detailed in \cite{FC} is based upon the measurement of the angles of impact of the drops in the pattern and on the estimated value of their uncertainties. The lines tangent to the trajectories at the points of impact are employed to construct a probabilistic map of the horizontal plane, related to the probability of convergence of their planar projections. The key idea is to probe the points in the horizontal plane in search of the areas of largest probability of intersection for the projections of the trajectories, given that their orientation can be associated to a certain statistical distribution, suggested by the error analysis.

To this end one must construct and calculate a joint probability density function (PDF), identified by the notation $\rho_{joint}{(P)}$, relative to the probability of convergence of all the projected trajectories at point $P$. This PDF is constructed by considering the uncertainties of the projections of the angles of impact as variances of a wrapped normal distribution (see the original article for the details).

The probability ${Prob}~{(\Sigma)}$ that a given area $\Sigma$ contains the horizontal projection of the point origin of the blood is then calculated by mean of an intergration of this PDF over $\Sigma$ itself:
\begin{equation}
{Prob}~{(\Sigma)}=\int_{P\in\Sigma}\rho_{joint}{(P)}{d\Sigma}.
\label{eq_Sigma}
\end{equation}
Even if a large number of tests would be necessary to prove the validity of this equation - and this is actually impossible given the time necessary to perform one whole experiment and the impossibility to reproduce analogous patterns to repeat the test under the same conditions - we show here that the outcomes of the method, which are essentially numerical, provide good agreement with the experimental measurements and with the reasonable expectations concerning the effect of variables like distance, angle of impact and number of stains in the pattern.

\section{Laboratory tests}\label{LT}
Some experiments have been performed in our laboratories to check the validity of the scheme presented in the previous paragraph. 
In this article we summarize two most significative examples, which are resemblant of typical forensic situations. Other examples, giving analogous results, are neglected for sake of shortness. In the following sections the used materials and the applied methods are preliminarily detailed.

\subsection{Materials}
The stage prepared for the experiments consist in two 4-wheeled wooden walls, 2 x 2 m of size, finished with smooth, white paint. Together with the floor, these two movable planes provide a flexible solution for simulating impacts from different distances and on surfaces oriented at different angles.
The fluid used to simulate blood has been prepared by uniformly mixing 80 cc of glycerol, 20 cc of distilled water and a few drops of water-soluble blue dye, to enhance stains visibility.

\begin{figure}
\centering
\includegraphics[width=0.9\textwidth]{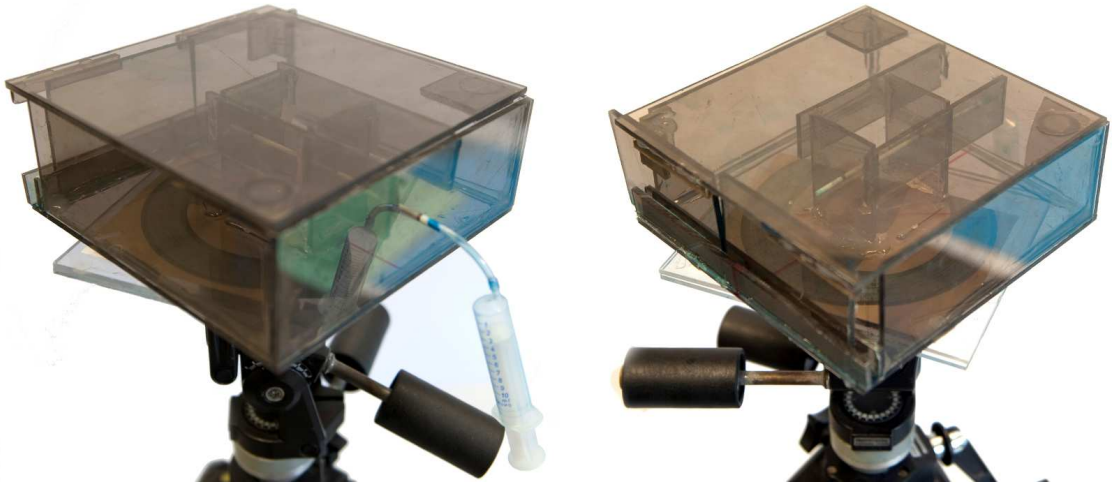}%
\caption{The spurting device used for the experimental tests. The spurting point is centered with the axis of the instrument and connected to the syringe containing the fluid by mean of a thin rubber tube. During the controlled spurting the direction of ejection can be varied without changing the coordinates of the origin. One thin slit provides a reduction of the azimuthal angle of emission and contribute to generate a more definite, less \emph{noisy} pattern.}
\label{fig:tool}
\end{figure}

The device employed for the spurting (see Fig.~\ref{fig:tool}) is a box, made of 5 mm-thick polycarbonate, with dimensions 15x15x5 cm; it has a removable top and two sliding and removable windows to adjust and limit the spraying angle. The liquid is drown into a 10 ml-syringe, external to the box and connected to the device by mean of a thin flexible rubber pipe, 15 cm long, whose other side is fixed and centered in the box itself. The instrument is also equipped with a fastening system that secures it to a tripod, allowing rotation along two independent axes. A spurting line is marked inside the box, for reference. This reference line, overlapping the photograph of a goniometer, glued on the bottom of the box, allows the measurement of the azimuthal spurting angle, with one degree of accuracy. With this setup the projection can be modulated in pressure without shaking the device, even during the rotation of the instrument itself, and the error on the measurement of the real position of the origin of trajectories is therefore minimized.

For every experimental setup, the height of the point of spray from the floor and the distances from the walls of the scenario have been measured with a flexible ruler; to adjust the angle between the planes of impact a mechanical-digital level and measurer Bosch\textsuperscript\textregistered~DWM~40L has been used.

The close-up photographs of the single selected stains in the patterns have been taken with a Nikon\textsuperscript\textregistered  D300 camera equipped with macro lens AF-S VR Micro Nikkor\textsuperscript\textregistered 105 mm. The camera was generally held by a Manfrotto\textsuperscript\textregistered~tripod, to minimize the blurs due to increased exposure times.

For digital photograph analysis we have used  Gimp 2.8 and Adobe\textsuperscript\textregistered~Photoshop\textsuperscript\textregistered~CS4, while for the calculations of the angles of impact of the drops and their errors and to calculate the area of convergence we have employed the BPA software AnTraGoS 2.4 of the Italian State Police (\emph{Polizia di Stato}). The sketches of the scene and the presentation of the results in the figures have been again realized and edited with AnTraGoS 2.4 and Gimp 2.8.

\subsection{Methods}
In this section we describe in detail two tests, \textbf{T1} and \textbf{T2}. The first test (\textbf{T1}) has been performed by spurting the liquid from a point equidistant from two orthogonal vertical surfaces (conventionally oriented along the $x$ and $z$ axes), at $66\pm 3$ cm of distance from both. During the experiments, the azimuthal angle of projection has been gradually varied, in order to produce an uniform deposit of blood on the walls in front. The fluid has been spurted from an height of $95\pm 3$ cm, with an upward elevation of about 45$^{\circ}$ (this value is however irrelevant). The error in the linear measurements has been estimated by considering the average movement of the source during the spurting operation.

Following the prescriptions of \cite{W}, some stains have been chosen on the impact surfaces, paying attention to (in order of importance): i) select well shaped, regular elliptical stains; ii) exclude satellite or secondary drops (generated by splitting of primary drops); iii) select drops quite distant from each other (to reduce redundancy); iv) select drops at different heights to provide uniform sampling on vertical axis; v) select stains of different sizes. Seventeen stains have been overall selected for the analysis, ten on the $z$-wall and seven on the $x$-wall; no stain has been selected from the floor.

Some detailed macro photographs have been taken for every single selected stain, including in the picture a vertical reference line\footnote{A plumb line has been obtained by mean of a thin fishing wire knotted to a flat steel disk; the flatness of disk is necessary to keep the line close to the wall and to produce focused photographs.} and a ruler.

The next step consisted in analysing and measuring the shapes of the photographed stains. These delicate operations have been repeated by three different operators making use of graphical softwares that allowed the close fitting of the outline of the stains with ellipses, and the measurement of the length of their axes. 

One suitable way to guess, for every stain, the most significative ellipse, and to estimate the related uncertainties in the measurement, has been the definition of two \textit{probe} ellipses, respectively the largest contained in and the smallest external to the stain itself. From the values of the length of the two couple of axes the best value and the statistical error have been directly estimated. For stains on vertical surfaces, the angle $\phi\pm\delta\phi$ between the major axis of the ellipse and the vertical line has been again graphically measured and then inserted in AnTraGoS together with the values of the axes of the ellipses, for the computation of the angles of impact and the relative calculated uncertainties.

The area of convergence has been calculated from the horizontal projections of the angles of impact of seventeen drops and from their statistical uncertainty, by mean of Eq.~\ref{eq_Sigma}. We have chosen a 1-cm spacing for the discretization in the numerical calculation of the integrals.

The second test (\textbf{T2}) has been performed from a non-symmetrical position: this time the source of the spurting has been located farther from one of the two vertical surfaces,  at $159\pm 3$ cm from the wall $x$ and  at $64\pm 3$ cm from the wall $z$. The height of the source has been placed at $150\pm 3$ cm and the elevation has been set to 0$^{\circ}$. Again, we have selected ten stains on the $z$-wall and seven stains on the $x$-wall. For the rest, the experimental and analytical procedure has been the same as in \textbf{T1}.

\subsection{Results}
\begin{figure}
\centering \resizebox*{9.95cm}{!}{\includegraphics{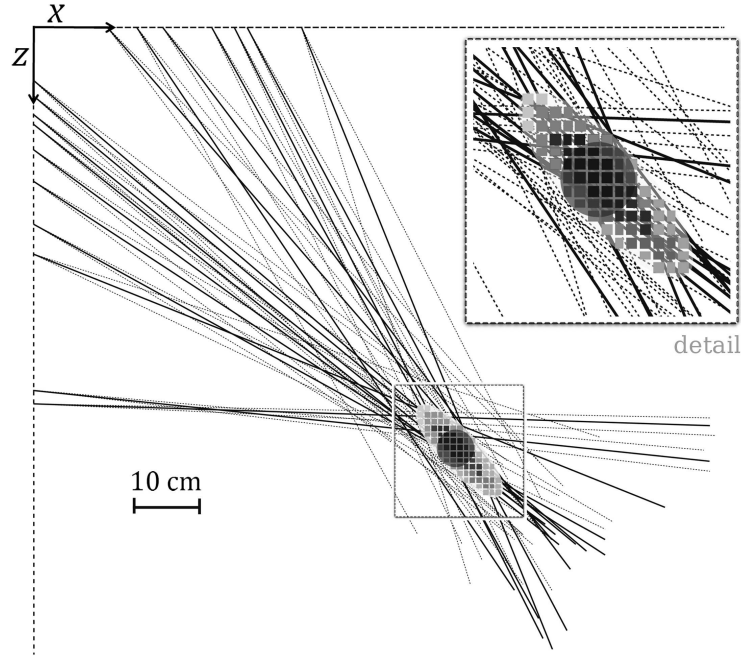}}
\caption{- \textbf{T1} - reconstruction of the area of convergence, performed with AnTraGoS. Decreasing levels of grey for the 1 cm-squares refer to levels of probability of convergence per cm$^2$ above $1.5\%$, between $0.5\%$ and $1.5\%$, between $0.1\%$ and $0.5\%$. The grey circle represents the area of actual origin of the projections. Being the results of the three experimenters very similar, only the result obtained by one of the three (i.e. \textit{E1}) is presented.}
\label{esp3}
\end{figure}

\subsubsection{General results}
The results of \textbf{T1} and \textbf{T2} are summarized in Table 1 and Table 2, respectively. The first three lines of the tables specify the values of the areas of convergence and the position of their centers\footnote{We define here the center of the area of convergence as follows: it is the point(s) of the horizontal plane where the joint probability of convergence $\rho_{joint}{(P)}$ has an absolute maximum.}, as independently calculated by the three experimenters. It is worth notice that the results for the areas of convergence largely overlap, even if the results for the single angles of impact slightly differed between the experimenters, following slight discrepancies on the measurement of the stains. This last fact is also responsible for the non-uniformity of the values of the areas. The second last lines report the result obtained by taking as an input for the calculation of the area of convergence the average value ($AVG$) of the angles of impact of the three experimenters, and the related statistical uncertainty.

\begin{center}
\begin{minipage}[b]{.65\linewidth}
\captionof{table}{Results of \textbf{T1} test}
\begin{tabular}{c c c}
\hline
ID  & Area of convergence & Center (\textit{x},\textit{z}) \\
Experimenter & (probability level: $95\%$)& [cm]\\
\hline
\textit{E1}  & 98 cm$^2$& (66, 68)\\ 
\textit{E2}  & 79 cm$^2$& (67, 69)\\ 
\textit{E3}  & 77 cm$^2$& (65, 65)\\ 
\hline
\textit{AVG} & 103 cm$^2$& (66, 68)\\
\hline\hline
\multicolumn{2}{c}{Actual area of origin} & (66$\pm$3, 66$\pm$3)\\
\label{T1}
\end{tabular}
\end{minipage}
\end{center}

\begin{center}
\begin{minipage}[b]{.65\linewidth}
\captionof{table}{Summary of \textbf{T2} test}
\begin{tabular}{c c c}
\hline
ID  & Area of convergence & Center (\textit{x},\textit{z}) \\
Experimenter & (probability level: $95\%$)& [cm]\\
\hline
\textit{E1}   & 135 cm$^2$& (64, 161)\\
\textit{E2}  & 95 cm$^2$& (62, 156)\\
\textit{E3} & 135 cm$^2$& (68, 165)\\
\hline
\textit{AVG} & 145 cm$^2$& (65, 162)\\
\hline\hline
\multicolumn{2}{c}{Actual area of origin} & (64$\pm$3, 159$\pm$3)\\
\end{tabular}
\end{minipage}
\end{center}

\begin{figure}
\centering \resizebox*{9.95cm}{!}{\includegraphics{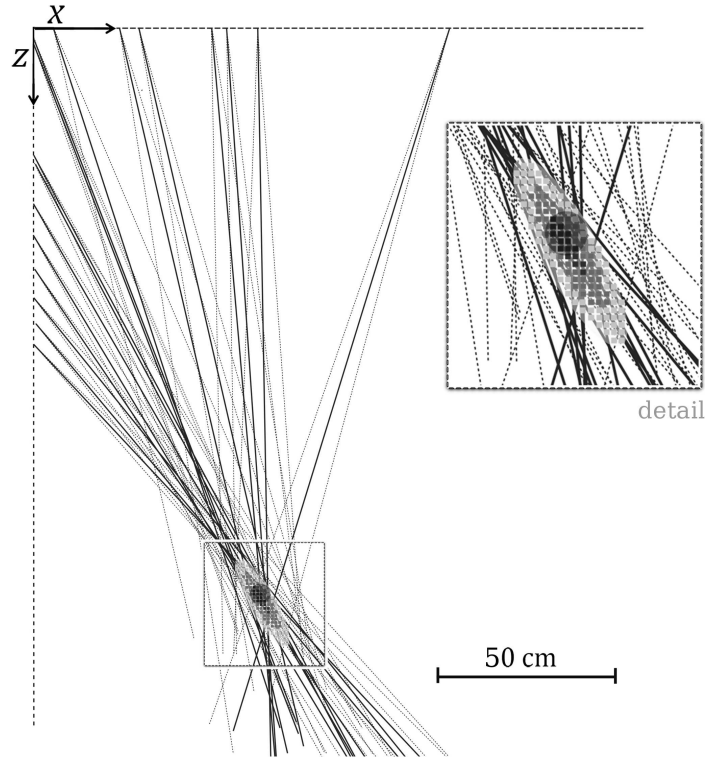}}
\caption{- \textbf{T2} - reconstruction of the area of convergence, performed with AnTraGoS. The same conventions as in Fig.~\ref{esp3} hold. Being the results of the three experimenters very similar, only the result obtained with the average of the three (i.e. \textit{AVG}) is presented. The area of convergence has an elliptical layout, oriented along the direction of the nearest stains.}\label{esp2}
\end{figure}

The three areas spotted by the experimenters are well overlapping the actual point of origin, shown in the last line of the tables. For reasons of clarity and shortness, being the graphical outcomes very similar, we decided to present, for each test, only one of the graphical outputs (e.g. \textit{E1} in Fig.~\ref{esp3} and \textit{AVG} in Fig.~\ref{esp2}).
The circular area of uncertainty, relative to the actual spurting point, is represented in grey in the figures. The area of convergence depicted in the figures is representative of $95\%$ probability of convergence, with darker areas relative to higher values of probability of convergence: decreasing levels of grey refer to levels of probability per cm$^2$ above $1.5\%$, between $0.5\%$ and $1.5\%$, between $0.1\%$ and $0.5\%$.
As clear from the graphical overlap of the colors, the result of the calculation is well reproducing the experimental condition, both in \textbf{T1} and \textbf{T2}.

\subsubsection{Further analyses}

\begin{figure}
\centering
\subfigure[all stains]{\resizebox*{2.65cm}{!}{\includegraphics{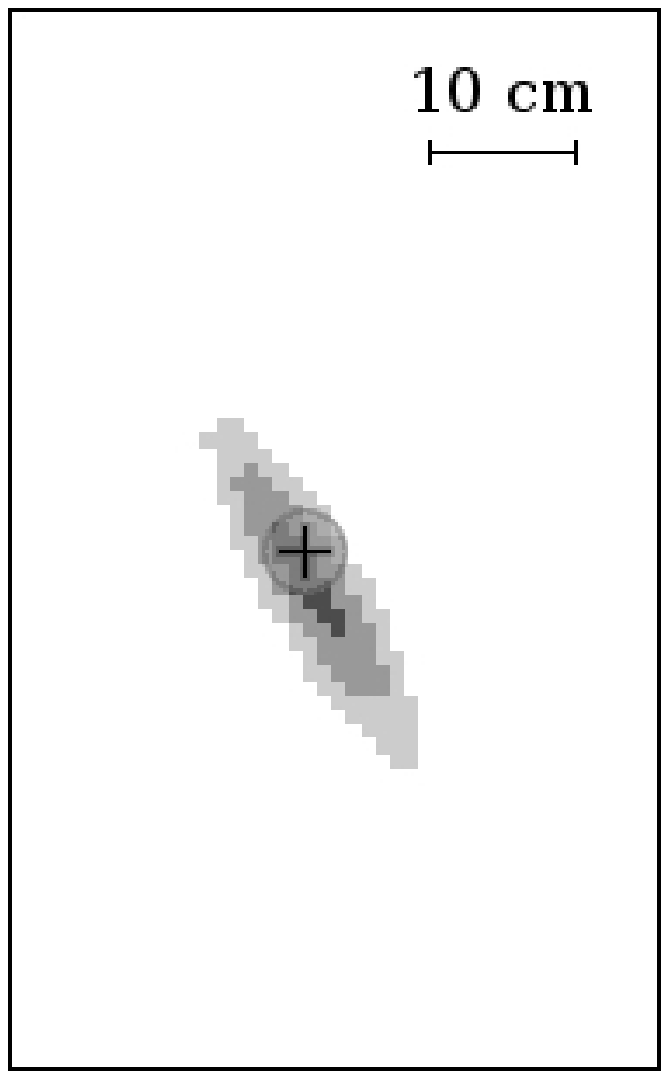}}\label{fig:i}}\goodgap
\subfigure[even stains]{\resizebox*{2.65cm}{!}{\includegraphics{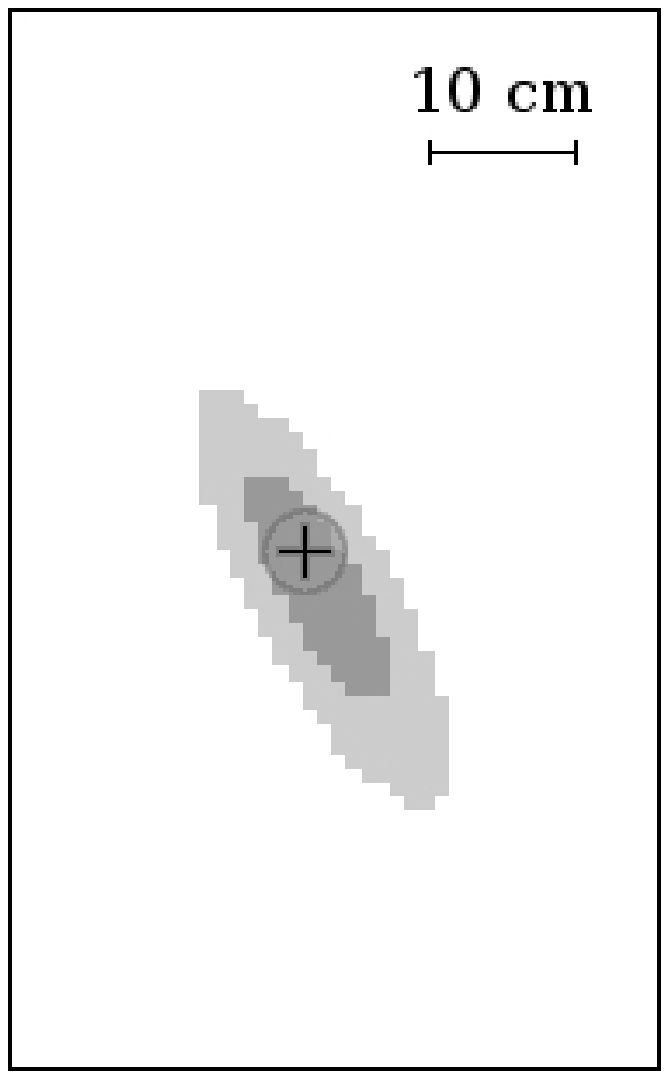}}\label{fig:ii}}\goodgap
\subfigure[odd stains]{\resizebox*{2.65cm}{!}{\includegraphics{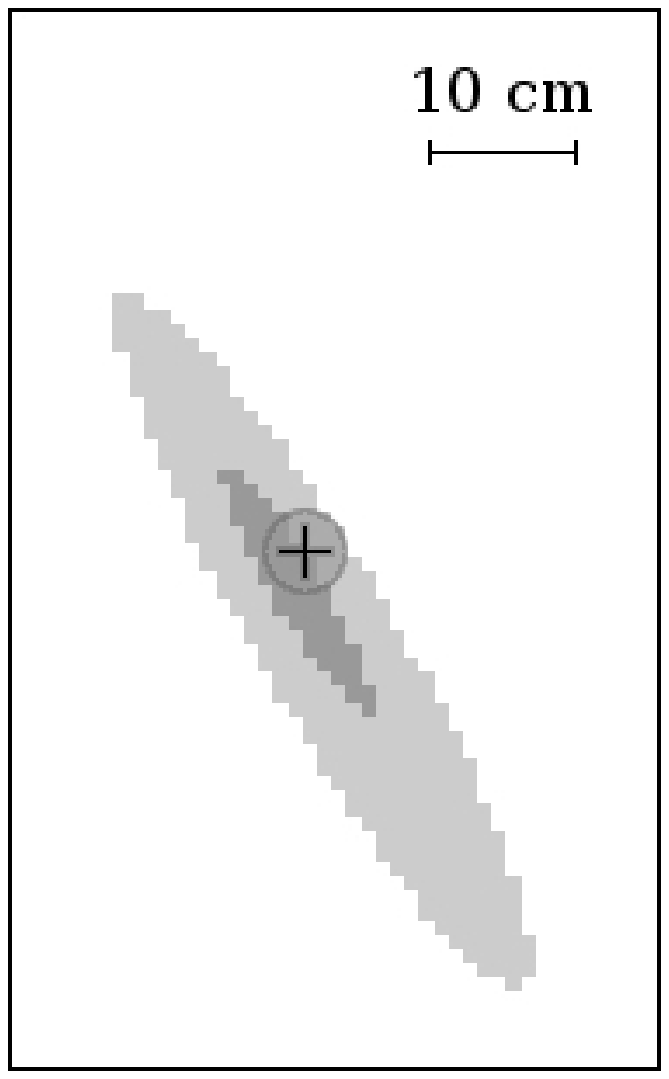}}\label{fig:iii}}\goodgap
\subfigure[$x$-wall stains]{\resizebox*{2.65cm}{!}{\includegraphics{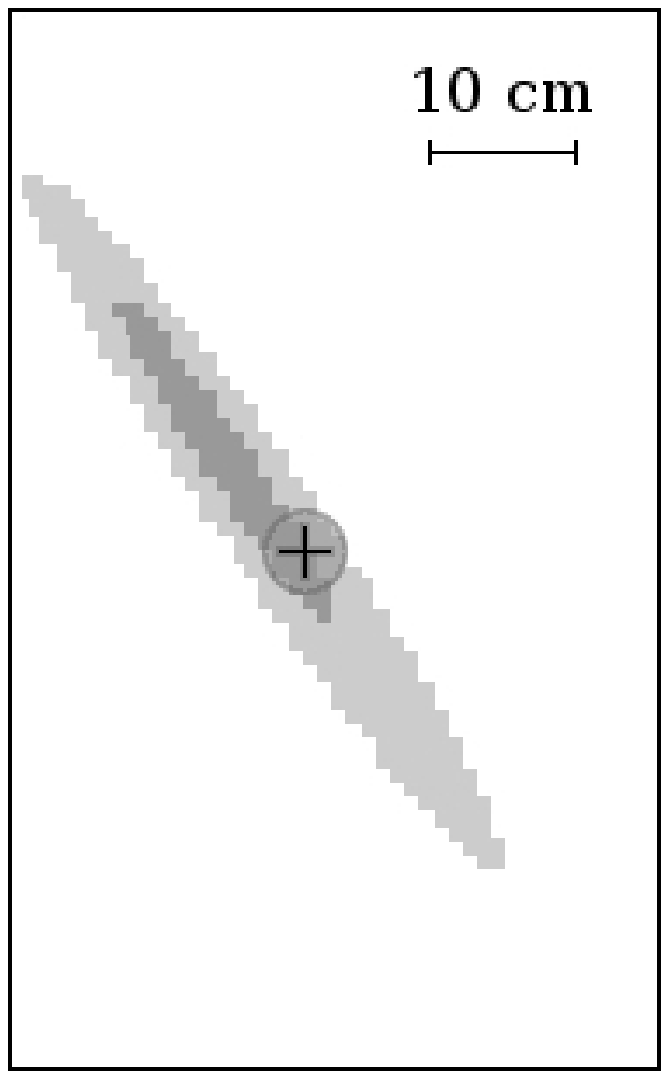}}\label{fig:iv}}\goodgap
\subfigure[$z$-wall stains]{\resizebox*{2.65cm}{!}{\includegraphics{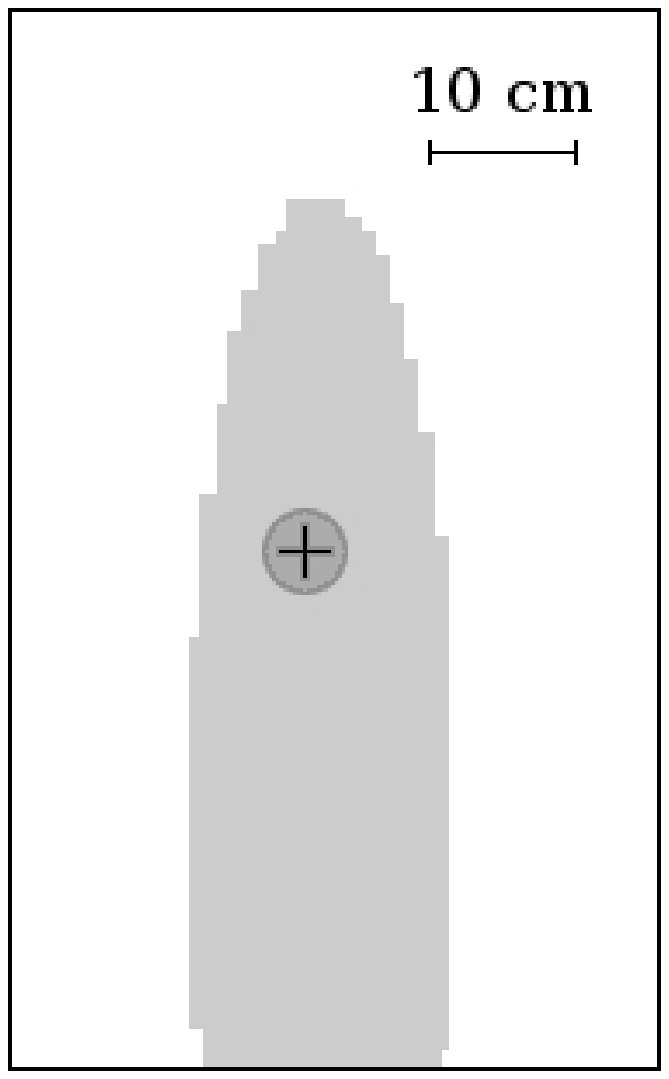}}\label{fig:v}}
\caption{\textbf{T2} test: selective analysis and reconstruction of the areas of convergence, performed with the inclusion of all the drops (a), of the even drops (b), of the odd drops (c), of the stains lying on the $x$ wall (d) and of the stains lying on the $z$ wall (e). The wider area in (e), compared to (d), is mainly due to the increased distance from the impact surface. The overall probability of the painted areas is always 95$\%$ (figure (e) has been cropped for uniformity of scale), while decreasing levels of grey refer to the same levels of probability detailed in Fig.~\ref{esp3}. The circular area represents the real spurting point.}
\label{fig:T2_2}
\end{figure}

In the previous section we have shown that the outcomes of all the three experimenters produce a good overlap between the actual projection of the area of origin and the calculated area of convergence. Both in \textbf{T1} and \textbf{T2} the result of the area of convergence obtained by averaging the three contributions (\textit{AVG}) appears to be more centered and widespread, compared to the single outcomes.

We decided to deepen the analysis of e.g. \textbf{T2}, to clarify some important details.
To do so, we have repeated the calculus on different subsets of stains, in order to probe the most relevant factors of the whole scheme (Fig.~\ref{fig:T2_2}).
The first step has been made by numbering the stains from 1 to 17, and by performing the analysis on the even and on the odd stains, separately. The results, depicted in Fig.~\ref{fig:ii} and \ref{fig:iii}, reveal wider areas of convergence with the same total probability of 95\%. 
This enlargement is due to the lower number of stains and in the fact that, being the lines quite well convergent, the statistical uncertainty decreases when the number of sampled stains increases.

Another important aspect concerns the distance between the point of origin and the point of impact. As easily confirmed by comparing Table 1 and 2, the value of the area of convergence increases with the distance: this is quite obvious being the errors of angular type. This aspect is also verified by comparing the results of Fig.~\ref{fig:iv} and Fig.~\ref{fig:v} showing the calculation preformed by separating the stains deposited on the $x$- and on the $z$-planes: actually the 95\% area of convergence  is very wide in this last case, both because of the distance of about 160 cm from the $z$-wall and for the lower number of stains (seven compared to ten). 

It is interesting to compare Fig.~\ref{fig:iv} and (e) with Fig.~\ref{fig:i}. This comparison highlights the importance of having stains on differently oriented walls, to produce a more definite result. If we imagine the area of convergence as an ellipse, its orientation follows the orientation of the trajectories: to reduce the elongation of this ellipse, it is recommended to have trajectories that converge in the area from different directions. And this is the reason why the inclusion of the $z$-wall stains of Fig.~\ref{fig:v}, even if distant, reduces significantly the uncertainty of the area  depicted in Fig.~\ref{fig:iv}, relative to the $x$-wall stains only, to produce the overall result of Fig.~\ref{fig:i}.

\section{Forensic applications}\label{FA}
The method described in \cite{FC} can be applied to a large set of forensic situations. In the following a simple applicative example is e.g. described, taken from an homicide scene, occurred in a province of northeastern Italy. 

A 52-year-old man was found dead at the entrance of his garage, after a dispute with his neighbor.
Its head presented a major wound, displaying a fracture of the right parietal surface of the skull.
A large quantity of blood, featuring a wide projective pattern, covered the two walls adjacent to the entrance and the floor of the garage, up to a maximum distance of approximately 3 meters. A pool of blood was extending from under the head of the victim.
Asked by the police, the neighbor asserted that a quarrel between him and the dead man had resulted in a violent contest at the end of which he had been punched in the face. Stunned by the punch, he had tried to push his offender back, who stumbled and fell to the ground, where he knocked his head and finally stood still, bloodied.

Another neighbor witnessed instead that he had heard and seen the quarrel, looking out of the window of his house, about 35 meters away, and, according to his testimony, the dispute terminated with a violent blow over the head of the victim who was trying to enter the garage. This blow with a blunt object had knocked the victim to the ground in the position where he had been ultimately discovered by the rescuers.
This version, so contrasting with that of the first neighbor, needed a precise reconstruction of the dynamics of the crime, which was obtained with the BPA.

\begin{figure}
\centering
\subfigure[Crime scene]{\resizebox*{7.65cm}{!}{\includegraphics{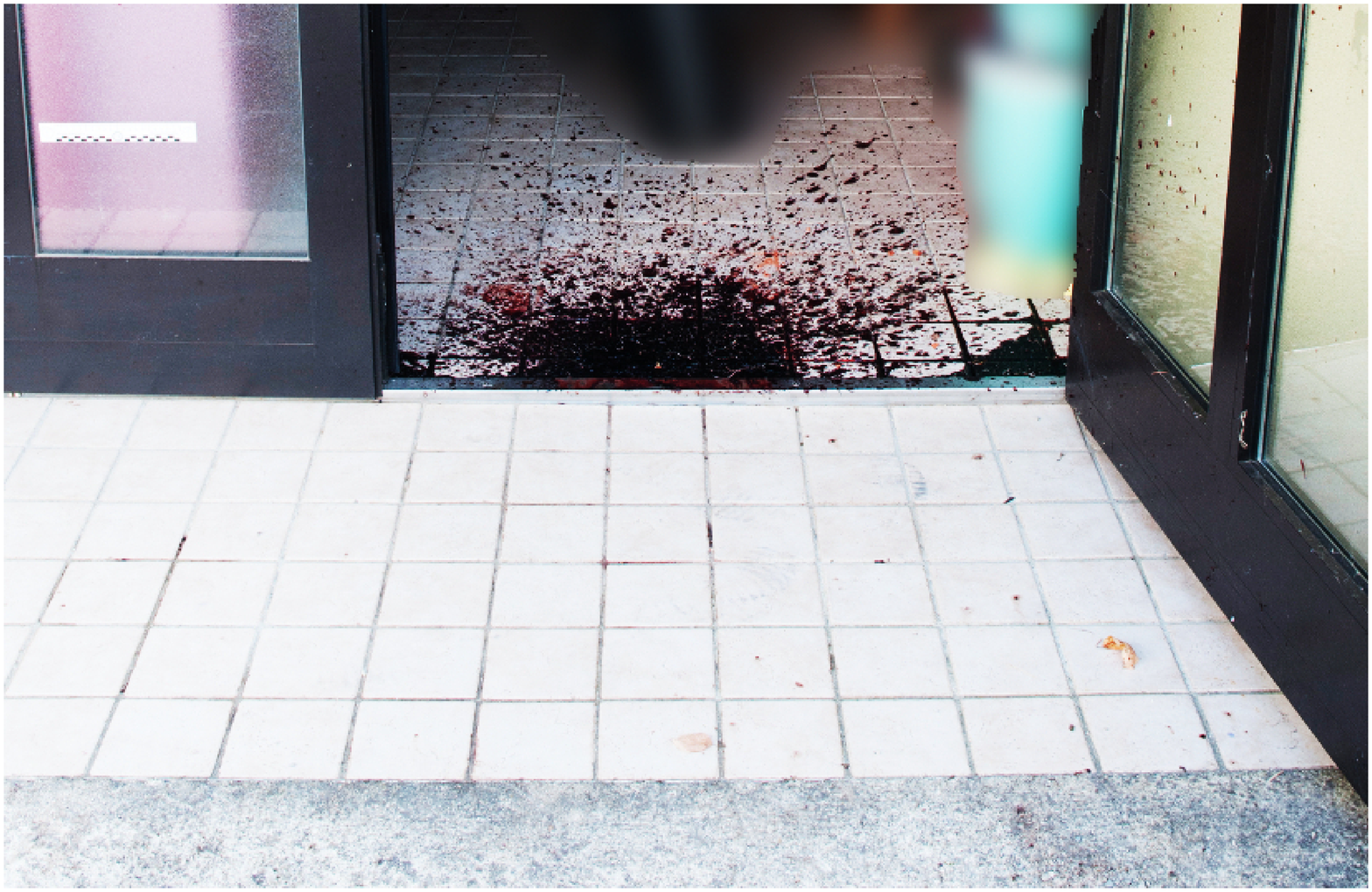}}}\goodgap
\subfigure[Areas of convergence]{\resizebox*{7.65cm}{!}{\includegraphics{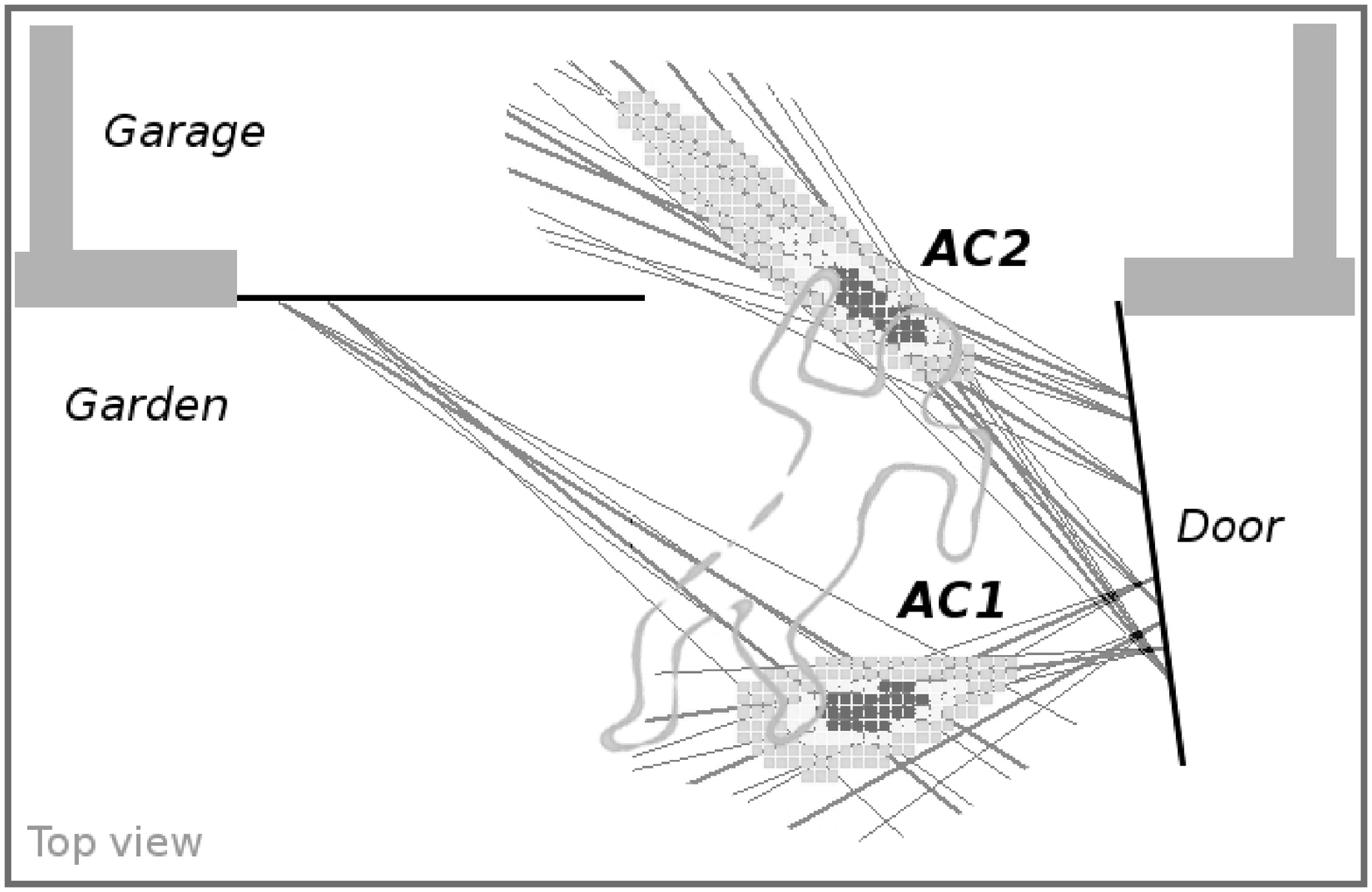}}}
\caption{An example of application of the probabilistic method to a real case (a). The reconstruction of the areas of convergence (b), performed with AnTraGoS, confirm that the action has occurred in two distinct phases. The impossibility to assign the blood projections to one single event (the fall of the victim) confirmed the incrimination of a neighbor, who insisted on a different, non-incriminating version.}
\label{real_case}
\end{figure}

The case was solved after a careful selection of the stains deposited on one open aluminium door of the garage.
At first glance, some of them seemed not to be compatible with an origin from the point of impact where the head of the victim hit the floor.
To determine their area of convergence a complete analysis was realized, according to the scheme described in the previous sections.
The result is presented in Fig.~\ref{real_case}.

It appears quite evident that, while some drops have a clear convergence at the point where the victim was found, some other trajectories aim to another point, about 1 m out of the garage. The size and shape of these stains, together with their reconstructed area of convergence, suggested an origin from an impact with a blunt instrument (spatter impact), or with its subsequent oscillation (cast-off pattern), but surely not with the impact of the man against the ground.

This thesis was supported by the subsequent detailed analysis of the single trajectories, and by the 3D-reconstruction realized with AnTraGoS to determine the area of origin of the drops, which confirmed the hypothesis of an height of origin of 140 cm or more (the same equations of motions as in \cite{VG,TW} have been used), completely contrasting the testimony of the first neighbor. The case was finally closed when the blunt object, a stone with the blood of the victim, was finally found abandoned nearby: this evidently confirmed the testimony of the second neighbor and incriminated the other.

\section{Conclusions}
In this technical paper we have applied the method presented in \cite{FC} to determine the area of convergence. We have described two of the tests performed to validate the analytical scheme, which are typical of commonly observed forensic situations. We could see that the results obtained by applying the procedure agree with the experimental measurements. Even if the statistical approach would have required a large number of tests to properly verify the whole mathematical scheme that drives the experimenter to the identification of the area of convergence, we could at least apply and test it in detail. Not only the area is verified to lie where it should - at the meeting point of the projections of the trajectories - but also its position, its orientation, its value are identified and characterized in mathematical, or better statistical terms. Moreover, the agreement between the results independently obtained by different operators confirmed the robustness of the procedure and the validity of the whole analysis process.

Compared to other methods, such as the one which identifies the area of convergence as the \textit{arithmetic mean} of the convergence points relative to every couple of projected trajectories \cite{IB}, this scheme generates more reliable results, because of the inclusion of an error analysis. The area of convergence is characterized and identified as an area of probability of intersection: the inclusion of the statistical error strengthens the value of the result, because it also considers the relevant degree of uncertainty related to the process of production of the blood pattern itself.

We have also presented a sample application to a real forensic case. We could demonstrate that the method is very flexible, in that the inclusion/exclusion of single sets of stains easily drive the forensic expert to the identification of the different areas where the action may have occurred. This is a necessary step to obtain the height of origin of the projections and to determine the dynamics of the crime.

The possibility to code the fundamental equations in a forensic BPA software largely simplified and shortened the work of the analysts. Even if an expert's eye is generally preferable for a reasonable interpretation of the data, the results can then also be read and visualized by non-expert users, which is often the case in a judicial context.

\section*{Acknowledgements}
We would like to acknowledge Roberto Mangione, Luca De Rosa and Diego Testolin for their effective support during and after the tests. We also acknowledge the Italian \emph{Polizia di Stato} for the granted permission.
\bibliographystyle{abbrv}


\end{document}